\documentclass[12pt]{article}
\usepackage{amssymb}
\usepackage{amsmath}
\usepackage{graphicx}
\usepackage{pazha}
\usepackage[round]{natbib}

\tightenlines

\sloppypar

\begin{document}
\baselineskip 21pt

\title{\bf ``Horseshoe'' Structures in the Debris Disks
of Planet-Hosting Binary Stars}

\author{\bf \hspace{-1.3cm}\copyright\, 2018 \ \
T. V. Demidova\affilmark{*}}

\affil{
{\it Central (Pulkovo) Astronomical Observatory, Russian Academy of Sciences, St. Petersburg, 196140 Russia}}

\vspace{2mm}

\sloppypar \vspace{2mm}

\noindent The formation of a planetary system from the protoplanetary disk leads to destruction of the latter; however, a debris disk can remain in the form of asteroids and cometary material. The motion of planets can cause the formation of coorbital structures from the debris disk matter. Previous calculations have shown that such a ring-like structure is more stable if there is a binary star in the center of the system, as opposed to a single star. To analyze the properties of the coorbital structure, we have calculated a grid of models of binary
star systems with a circumbinary planet moving in a planetesimal disk. The calculations are performed considering circular orbits of the stars and the planet; the mass and position of the planet, as well as the mass ratio of the stars, are varied. The analysis of the models shows that the width of the coorbital ring and its stability significantly depend on the initial parameters of the problem. Additionally, the empirical dependences of the width of the coorbital structure on the parameters of the system have been obtained, and the parameters of the models with the most stable coorbital structures have been determined. The results of the present study can be used for the search of planets around binary stars with debris disks. 

\noindent Keywords: {\it modeling, planetesimal disks, binary star systems, disk and planet interaction}.

\vfill
\noindent\rule{8cm}{1pt}\\
{$^*$ E-mail: $<$proxima1@list.ru$>$}

\section*{INTRODUCTION}

More than $70\%$ of main-sequence stars are members of binary or multiple star systems. The studies of star-forming regions have shown that binarity is also common among young stars~\citep[see, e.g.,][]{1999A&A...341..547D,2009ApJ...704..531K}. This fact supports the scenario of simultaneous formation of binary and multiple systems from a protoplanetary cloud~\citep{1994MNRAS.271..999B}. The numerical calculations
have shown that circumstellar disks around each component of the system, as well as a common circumbinary disk, can form at the stage of accretion of matter from the remains of the protoplanetary cloud \citep{1996ApJ...467L..77A,2010ApJ...708..485H,2010ARep...54.1078K,2013A&A...556A.148P}. The circumstellar protoplanetary disks were discovered in
the observations of wide pairs~\citep{2010Sci...327..306M}, while the circumbinary disks were observed in close binary systems~\citep{2010ApJ...720.1684R,2011A&A...528A..81P}. The formation of large-scale structures in the gas-and-dust disk, which can also be identified in observations, is the consequence of the orbital motion of the binary system~\citep{2015A&A...579A.110R}.

The further evolution of the protoplanetary disk leads to depletion of the gaseous component. However, unlike the case of a single star, the debris disk of the binary system undergoes the dynamic mixing of planetesimals, which can hinder the formation of planets \citep{2004ApJ...609.1065M,2012ApJ...752...71M,2012ApJ...761L...7M,2012ApJ...754L..16P}. \citet{2015ApJ...805...38D} showed that the orbital precession of planetesimals can lead to the formation of a spiral structure in the circumbinary disk. Depending on the parameters of the binary system, the time of spreading of this structure across the disk can exceed the lifetime of the gaseous component and characteristic time of formation of planetesimals. A gap is formed in the center of the circumbinary disk, matching the zone of chaotic motion caused by overlapping of the orbital resonances of planetesimals with the central binary star~\citep{2015ApJ...799....8S}.

The analysis of the observable debris disks has shown that only a quarter of those belong to binary stars~\citep{2015MNRAS.449.3160R}. This effect may be associated with the fact that the gravitational influence of the binary system can accelerate the destruction of the debris disk due to frequent collisions between planetesimals~\citep{2015MNRAS.449.3160R}. \citet{2007ApJ...658.1289T} studied the spectral energy distribution of main-sequence binary systems of spectral classes A3-F8 with an age above $600$ Myr. It was shown that a third of the observed stars had an infrared excess, conditioned by the thermal radiation of dust that formed due to collision of planetesimals. The IR excess of the binary stars was higher than that of the single stars of the same age in spectral classes AFGK, especially in the case of close systems ($<3$~AU). These results indicate that the planetesimal disks of binary systems are more disturbed than the single-star disks. A detailed study of two debris disks in close binary systems $\alpha$ CrB and $\beta$ Tri are presented in~\citet{2012MNRAS.426.2115K}.

The planets can already exist at the stage of the planetesimal disk, since they are formed in a gas-rich medium \citep[see, e.g.,][]{2012RAA....12.1081Z}. \citet{2016MNRAS.463L..22D} studied the influence of a circumbinary planet on the debris disk of a binary star. It was shown that a resonance coorbital circular structure formed along the planetary orbit. Similar structures were described in~\citet{2000ApJ...537L.147O,2003ApJ...588.1110K} for the case of a single central star.
However, in the present study, the evidence was obtained that the coorbital structure is more stable and massive in the binary star system.

A number of protoplanetary disks reveal bright ring-like structures, conditioned by the thermal radiation of dust. Such rings may be associated with the protoplanets forming in the disk. For example, in the
case of the protoplanetary disk of HD 169142, the existence of the ring-like structure can be explained by the presence of at least two planets that cleared the gap within and outside the ring \citep{2013ApJ...766L...2Q,2014ApJ...791L..36O}. In the case of the young star HL Tau, the ring-like structure can be associated with the forming planet~\citep{2016ApJ...821L..16C}. That being said, it was shown in~\citet{2016MNRAS.463L..22D}  that the disk structure, similar to that of HL Tau with alternating dark and bright ring-like bands, can be formed by the motion of a single planet. 

The question of stability of planetesimals in the debris disk of the planet-hosting binary star was considered in~\citet{2006MNRAS.368.1599V}  by the example of the system $\gamma$ Cep. The simulations have shown that the regions of stability can exist around each star, as well as within the circumbinary disk. The same authors also studied the region of stability for the planet around a triple star and dynamics of planetesimals in the system HD 98800, which consists of four stars~\citep{2007MNRAS.382.1432V,2008MNRAS.390.1377V}.

The resonance structures coorbital with planets were discovered in the Solar System as well. Such structures include the Trojan asteroids coorbital with Jupiter; similar objects were also discovered near other
planets \citep{1990BAAS...22.1357B,2006Sci...313..511S,2011Natur.475..481C,2013Sci...341..994A}. 

In the present work, we study the secular evolution of binary star systems with a circumbinary planet that moves in the planetesimal disk for a wide set of initial model parameters.

\section*{MODEL AND METHOD}

The model of the system in consideration consists of two stars with masses $M_1$ and $M_2$ and a planet with a mass $m_p$; all the three bodies at the initial time point are placed into coplanar circular orbits. The period of the binary star $P_b = 0.2$~years; the period of the planet $P_p$ varies and corresponds to the following $P_p/P_b$ resonances: $5/1$, $11/2$, $6/1$, $13/2$, $7/1$, $15/2$, and $8/1$. The
mass of the main component for all models $M_1 = M_\odot$; the mass of the secondary component $M_2$ ranges within $0.05-1M_\odot$; and the mass of the planet ranges between $0.1$ and $10$ masses of Jupiter. Additionally, for the comparative analysis, we have calculated a set of
models of a single central star with a mass $M = 1.2M_\odot$ and a planet at the same distance as the planet of a binary star, the planet mass varies within the same limits. 

The circumbinary disk of the system is formed by $20000$ massless planetesimals placed into initially circular orbits at a radial distance between $1$ and $15$ semimajor axes of the binary star ($a_b$); they are distributed radially in accordance with the law $\Sigma\sim R^{-1}$ surface density depending on the distance) and uniformly in the vertical direction from $-a_b$ to $a_b$. The self-gravity of the disk is negligibly small, since its mass is insignificant in comparison with the mass of the planet and binary star. The motion of planetesimals in the system is considered within the restricted four-body problem ``binary star--planet--planetesimal'': the planet disturbs the motion of the planetesimal, but does not affect the orbit of the binary star, while the planetesimal does not affect the motion of the massive bodies. The orbits of the binary system and the planet are calculated within the restricted three-body problem, i.e., the planet is considered to be a passively gravitating particle in the gravity field of the binary star. In this
case, the motion equation of the planetesimal is written as follows:
\begin{equation}
\frac{d\vec{v}}{dt}=\nabla\phi_1+\nabla\phi_2+\nabla\phi_p
\label{pot}
\end{equation}
where $\phi_1, \phi_2$ and $\phi_p$ are the gravitational potentials of
the binary components and the planet, respectively. Equations~(\ref{pot}) are integrated using the symplectic algorithm~\citep{1967PhRv..159...98V}; the accuracy of this method is $O(\Delta t^4)$. Choosing the small step of integration allowed setting the maximum time of calculations as $5 \times 10^4$ years. The selective backward integration in time for several particles has shown a close agreement with the initial data, which allowed us to control the accuracy of the calculations.

\section*{RESULTS}
\subsection*{Dynamics of Matter in the Coorbital Structure}

\begin{figure}
\centering \includegraphics[width=0.8\textwidth]{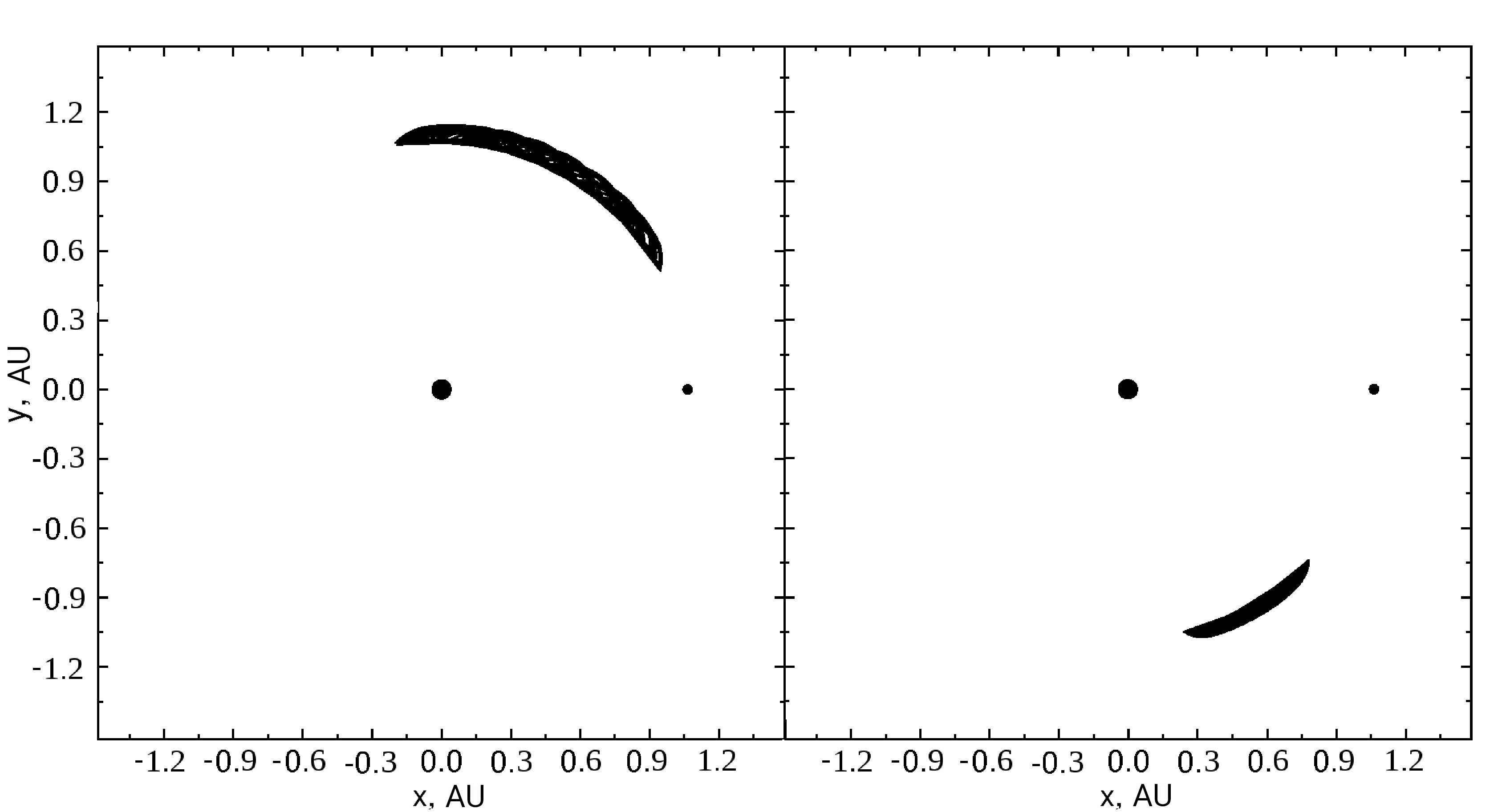}
\caption{Types of planetesimal orbits in the ring coorbital with the planet for the model with a single star in the reference frame
moving with the planet.} \label{fig1}
\end{figure}

\begin{figure}
\centering \includegraphics[width=0.8\textwidth]{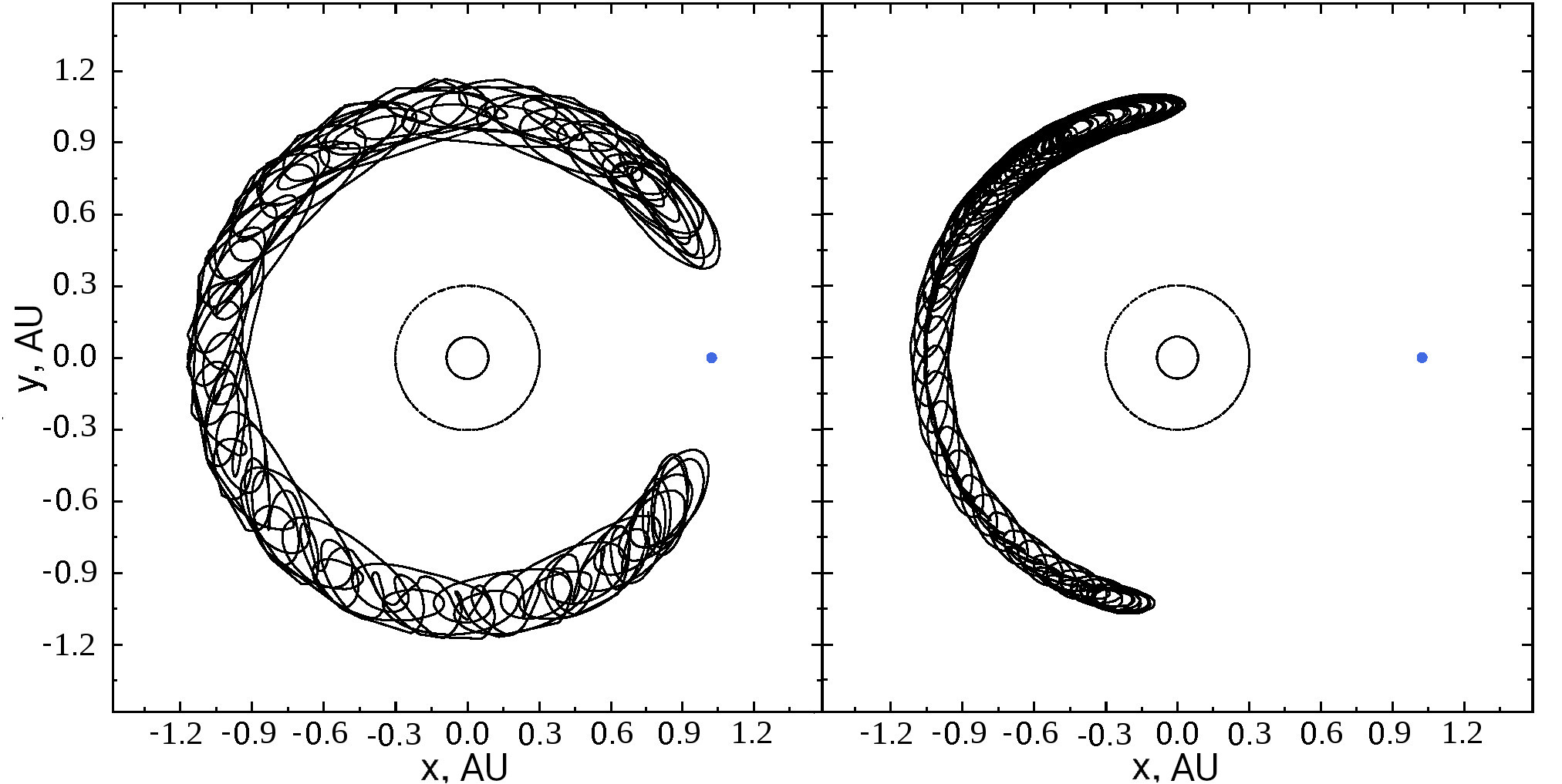}
\caption{The same as Fig.~\ref{fig1} for the binary star system.}
\label{fig2}
\end{figure}

The comparison of the character of planetesimal motion in the gravity field for a single and binary star shows that in the first case, the particles are captured into resonances at two Lagrange points, $L4$ and $L5$. In this case, two types of the planetesimal orbits in the coorbital ring can occur: ``tadpole'' (Fig.~\ref{fig1}) and ``horseshoe'' (Fig.~\ref{fig2}). The planetesimals in the single-star models prefer the first type of the orbits four times more often than the second type. However, the situation changes when there is a binary star in the center of the system: the planetesimals in the coorbital ring follow the orbits of the ``horseshoe'' type only (Fig.~\ref{fig2}), and the velocity in the ring relative to the planet decreases approximately by a factor of three in comparison with the single-star models.	

\begin{figure}
\centering \includegraphics[width=0.8\textwidth]{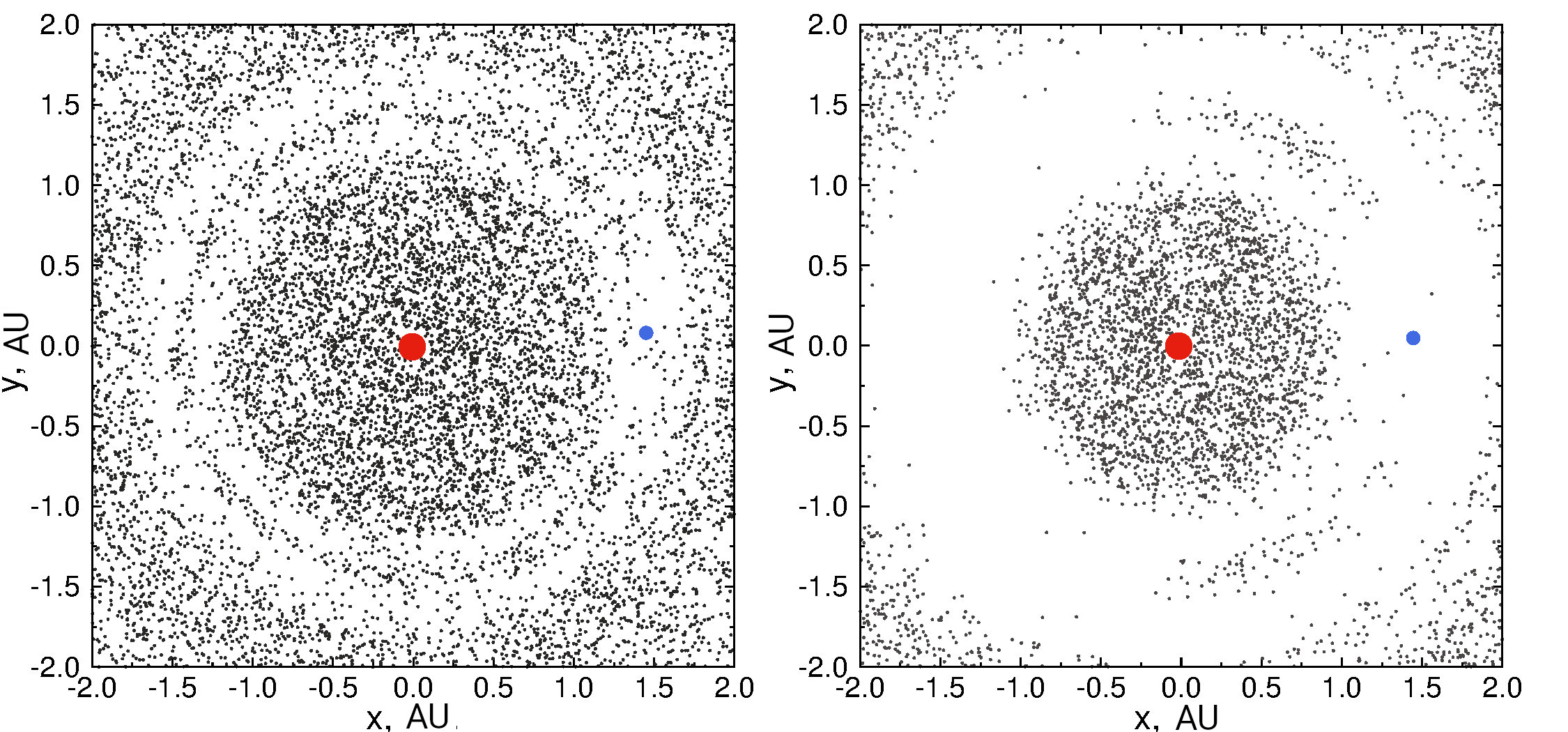}
\caption{Distribution of the debris disk material in the model of a single star with a planet at the time point of $5 \times 10^4$ years. The
mass of the planet is $0.6M_J$ (left) and $10M_J$ (right).} \label{fig3}
\end{figure}

\begin{figure}
\centering \includegraphics[width=0.8\textwidth]{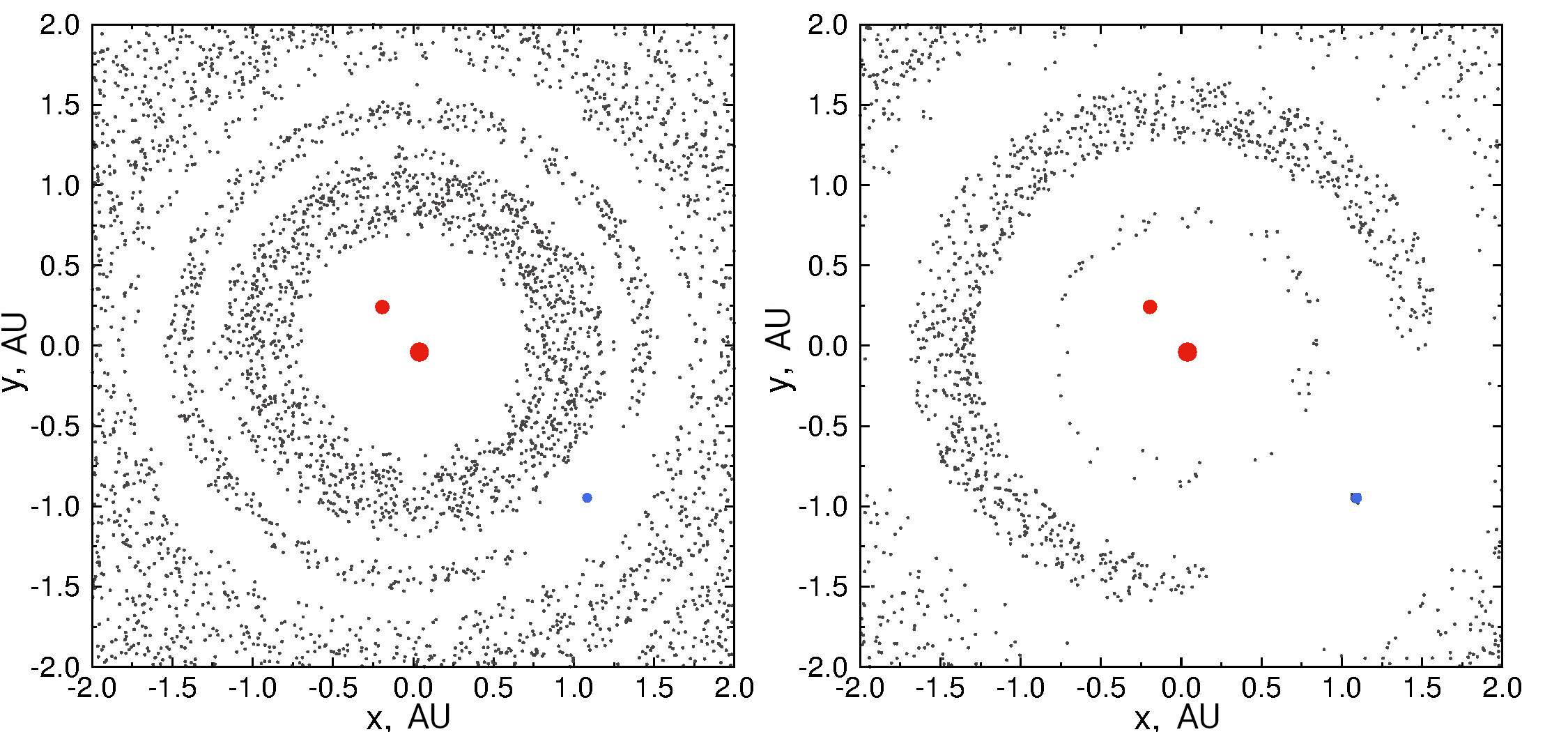}
\caption{The same as Fig.~\ref{fig3} for the binary star system.}
\label{fig4}
\end{figure}

Additionally, an increasing mass of the planet around a single star leads to a decrease in the number of planetesimals moving in ``horseshoe'' orbits (Fig.~\ref{fig3}), while the matter is concentrated toward the
Lagrange point. In the case of a central binary star, such an effect is not observed: with the increasing mass of the planet, the width of the ring increases as well (Fig.~\ref{fig4}).

\subsection*{Influence of the Model Parameters on the Width
of the Coorbital Structure}

To determine the width of the coorbital structure, we constructed the radial profiles of the surface density averaged over azimuth. The disk was divided into separate rings with a radius step of 0.002 AU. Within
each ring, the number of particles divided by the ring's area was calculated. The calculations showed that the characteristics of the coorbital structure over the period from $10^4$ to $5 \times 10^4$~years changed only slightly; for this reason, to eliminate random fluctuations, the surface density profiles were averaged on this time interval. The outer and inner boundaries of the coorbital ring were specified as the distances from the radial position of the planet, at which the difference of surface densities in the adjacent rings was $1\%$ of the surface density in the center of the coorbital ring. Thus, the distance was determined at which the radial profile of the surface density reached a plateau. The analysis of the results showed that such an approach
allowed a reliable determination of the size of the coorbital ring.

\begin{figure}
\centering \includegraphics[width=0.8\textwidth]{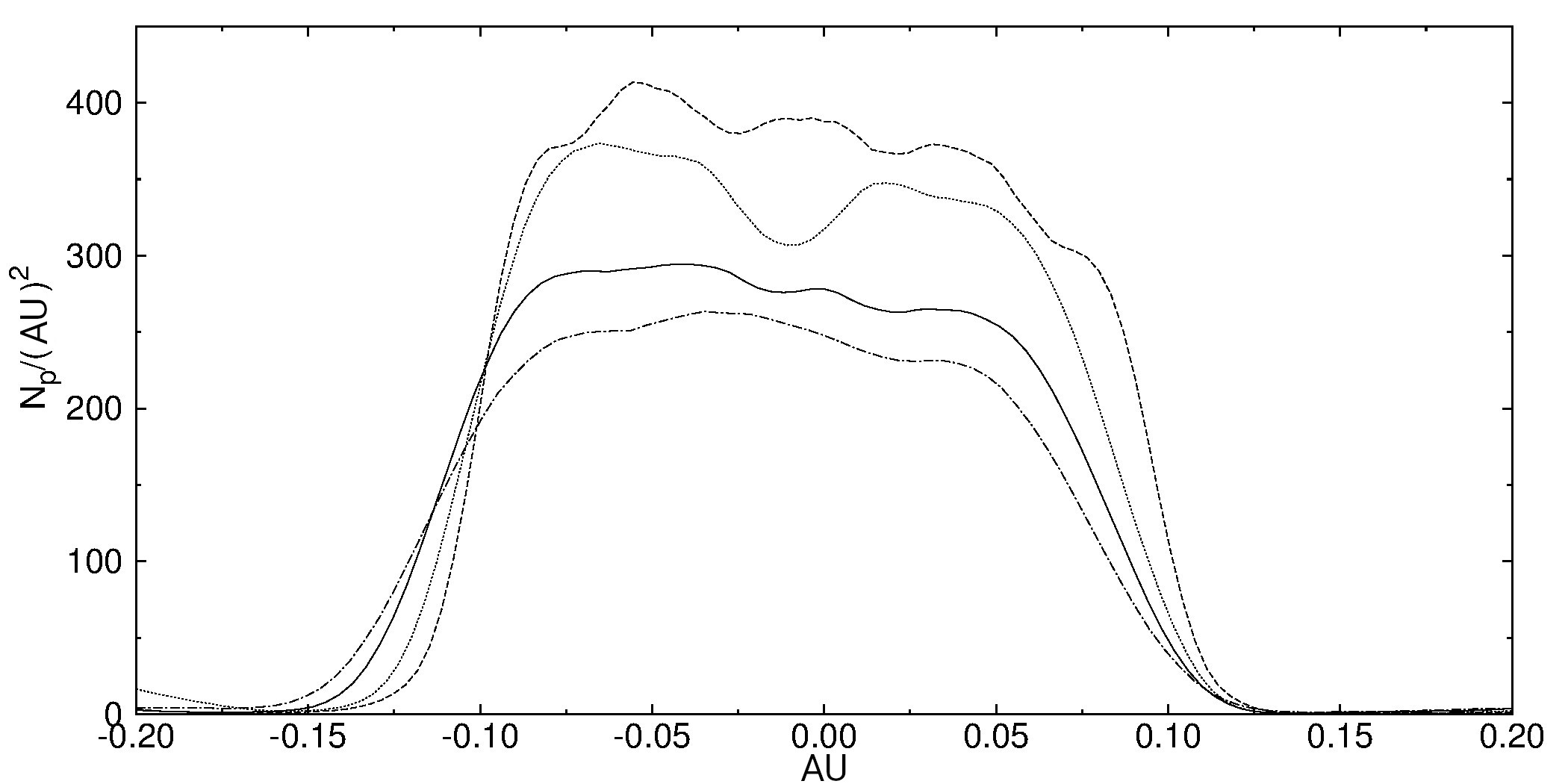}
\caption{The azimuth-averaged radial barycentric profiles of the surface density in units of number of planetesimals/(AU)$^2$ as a function of a radial position of the planet. For all models, the mass of the main component of the binary star $M_1 = M_\odot$ . The dashed line corresponds to the model with mass of the secondary component $M_2 = 0.05M_\odot$, dotted line corresponds to $M_2 = 0.2M_\odot$, solid line corresponds to $M_2 = 0.5M_\odot$, and dash-dotted line corresponds to $M_2 = M_\odot$.} \label{fig5}
\end{figure}

The comparison of the radial profiles of surface density for different models has shown that an increase in the mass ratio of the binary star components $M_2$ ($\mu_b=\frac{M_2}{M_1+M_2}$), given other equal parameters, leads to a shift in the maximum surface density from the
planet’s position toward the center of the system (Fig.~\ref{fig5}). Since the mass ratio of the planet and stars $m_p$ ($\mu_p=\frac{m_p}{M_1+M_2+m_p}$) decreases with increasing $\mu_b$, the surface density in the ring structure also declines.

\begin{figure}
\centering \includegraphics[width=0.8\textwidth]{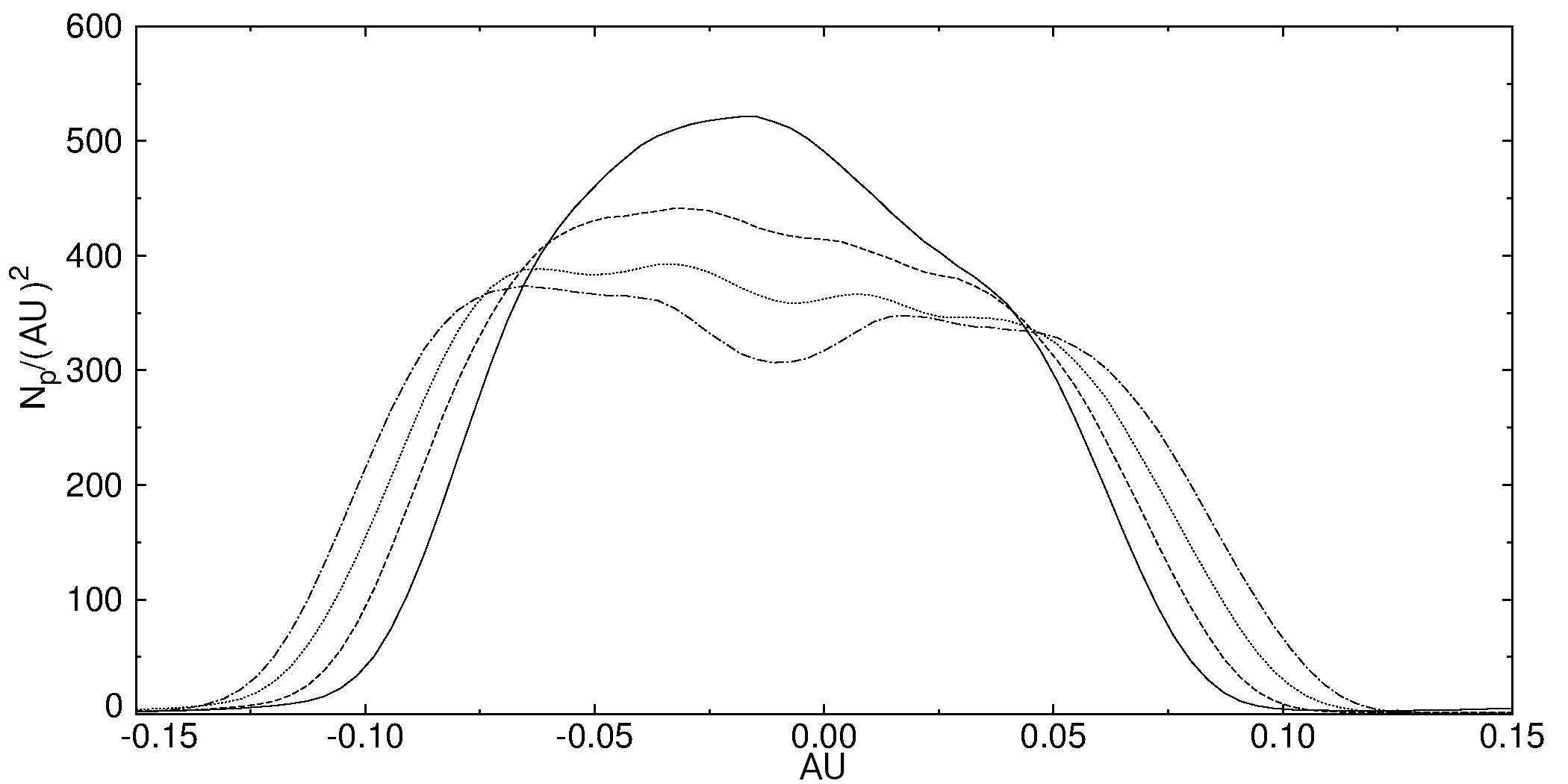}
\caption{The same as Fig.~\ref{fig5} for the model $\mu_b = 0.167$, $m_p = 10M_J$. The solid line corresponds to the position of the planet in the
$5:1$ resonance with the binary star, dashed line corresponds to $6:1$, dotted line corresponds to $7:1$, and dash-dotted line corresponds to $8:1$.} \label{fig6}
\end{figure}

\begin{figure}
\centering \includegraphics[width=0.8\textwidth]{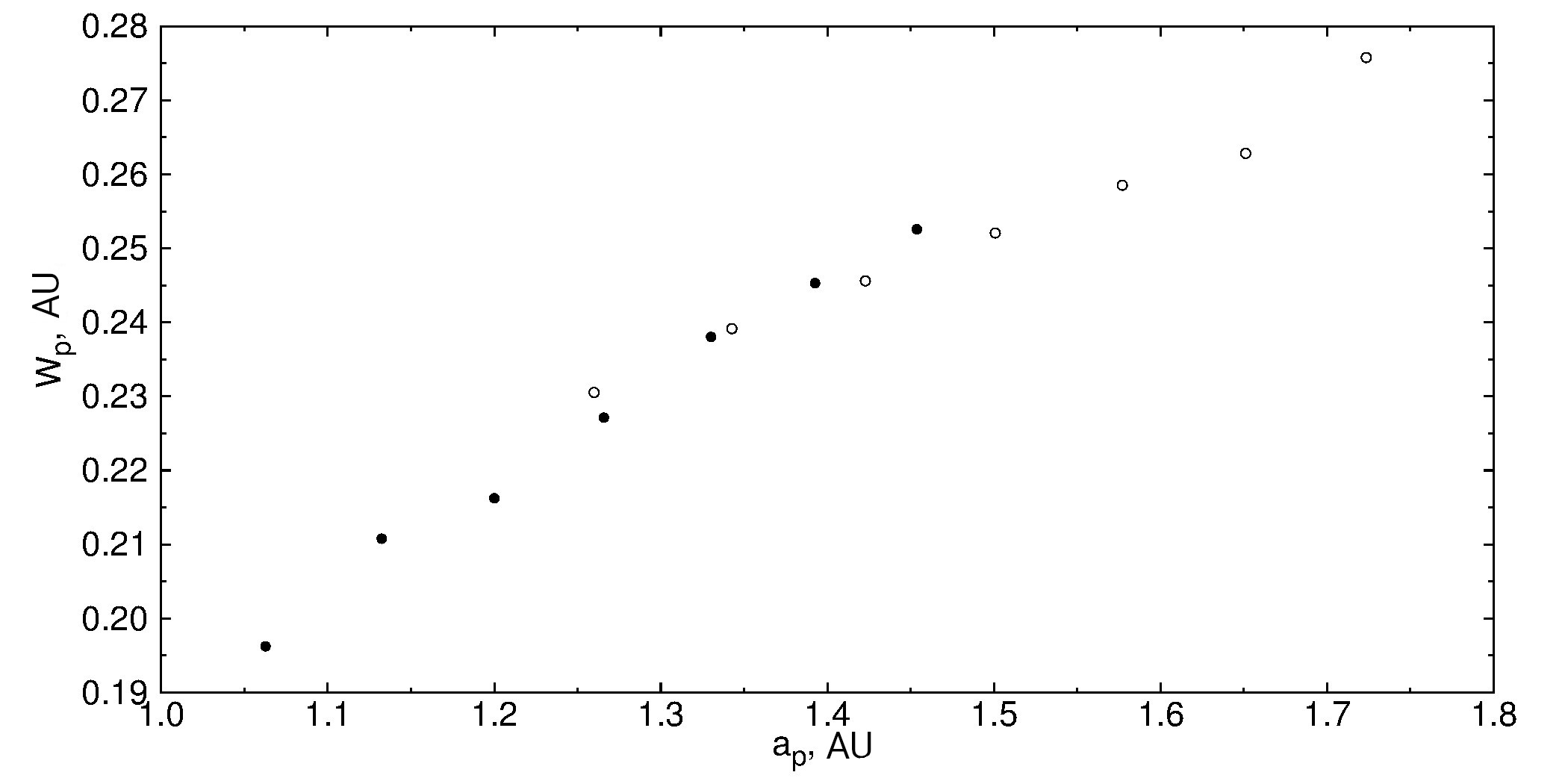}
\caption{The width of the coorbital ring as a function of the radial position of the planet. For all models, the mass of the planet $m_p = M_J$. The dark circles correspond to the model with the mass ratio of the binary star $\mu_b = 0.167$; white circles correspond to $\mu_b = 0.5$.}
\label{fig7}
\end{figure}

\begin{figure}
\centering \includegraphics[width=0.8\textwidth]{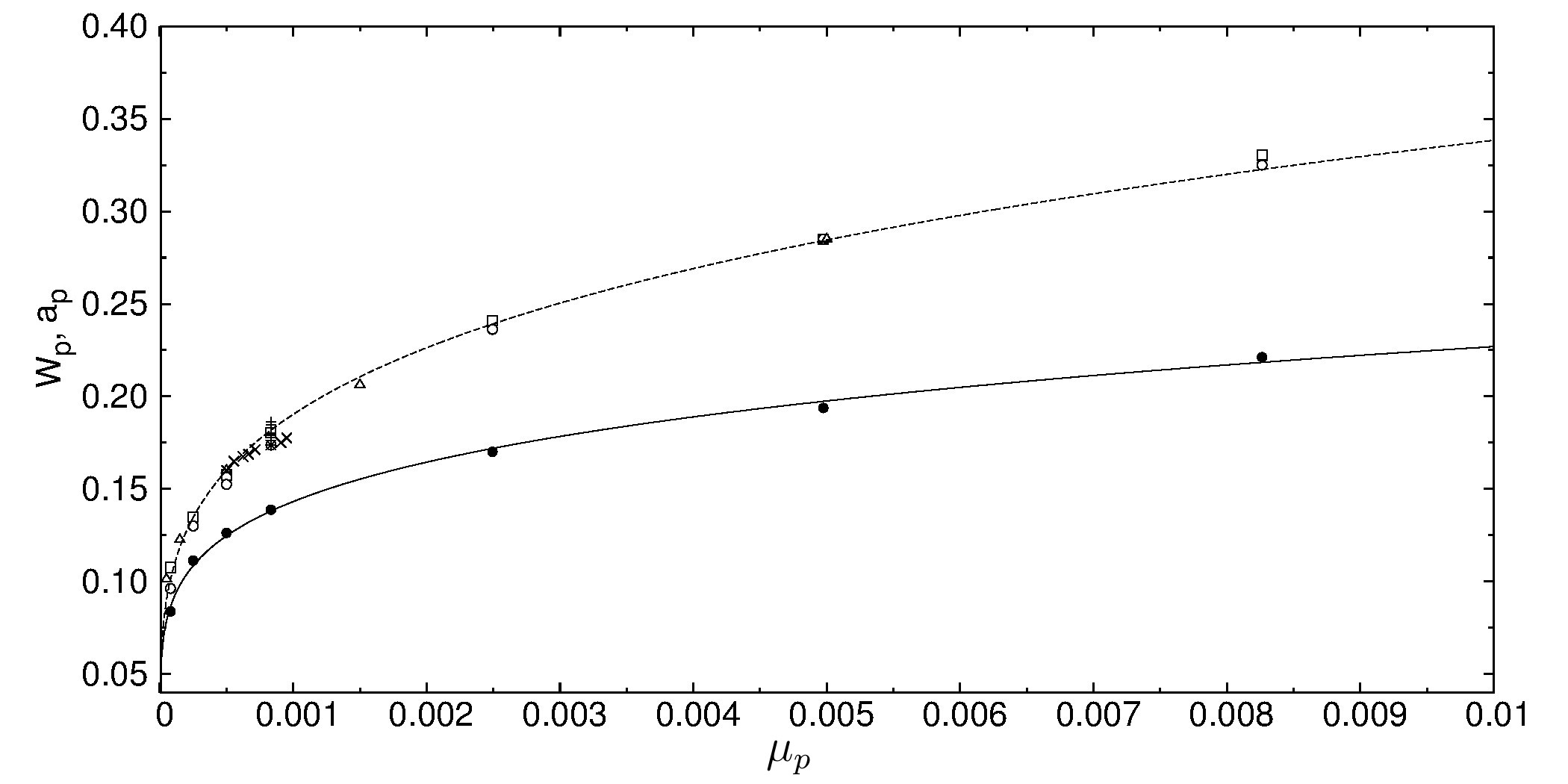}
\caption{The width of the coorbital ring normalized by $a_p$ as a function of the mass ratio of the planet and the stars. The symbol ``$\odot$'' corresponds to the models with $P_p/P_b = 6/1$, $\mu_b = 0.167$; ``$\circ$'' corresponds to $P_p/P_b = 8/1$, $\mu_b = 0.167$; and ``$\Delta$'' corresponds to $P_p/P_b = 8/1$, $\mu_b = 0.5$; the mass of the planet is varied. The symbol ``$\times$ corresponds to the models with the varying mass ratio of the binary star ($\mu_b$), $P_p/P_b = 8/1$, and $m_p = M_J$; ``$+$'' corresponds to $m_p = M_J$, $\mu_b = 0.167$, and varying $Pp/Pb$.}
\label{fig8}
\end{figure}

The model with the mass ratio of the binary star $\mu_b = 0.167$ features a distinct decrease in the density at the center of the coorbital ring near the planetary orbit. The same effect occurs for the model with $\mu_b = 0.09$; however, it is absent for other mass ratios of the components. When the planet's orbit approaches the binary star, the central trough of density disappears for these models as well (Fig.~\ref{fig6}). The calculations have shown that the width of the
coorbital ring, $W_c$, depends on the radial position of the planet, $a_p$, (Fig.~\ref{fig7}) and the mass ratio of the planet and the star, $\mu_p$ (Fig.~\ref{fig8}). Based on the analysis of $27$ models, an empiric dependence of $W_c$ on these parameters was acquired:
\begin{equation}
W_c \propto a_p\mu^{1/4}_p.
\end{equation}

In Fig.~\ref{fig8} this dependence is shown with the dashed line. The width of the coorbital structure for the case of a single central star shows a weaker dependence $W_c \propto a_p\mu^{1/5}_p$ (solid line in Fig.~\ref{fig8}). The dependence of the width of the coorbital structure on the mass ratio of the binary star is not found.
\newpage
\section*{Stability of the Structure}

To analyze the stability of the coorbital ring, the time dependence of the number of planetesimals within its boundaries was calculated with a step of $100$ years. Further, by comparing the number of particles in the ring at the final ($5 \times 10^4$ years) and initial ($10^4$ years) time points, the character of change in the amount of matter in the coorbital structure was studied. The calculations have shown that for most models,
the particles leave the coorbital structure. For the models with $\mu_p > 10^{–3}$, the percentage of planetesimals remaining at the final time point is the smallest and amounts to less than $98\%$. If $P_p/P_b < 6/1$, such a behavior is also true for the models with $\mu_p > 8 \times 10^{–4}$. However, for the models with $\mu_p < 7 \times 10^{–4}$ and $P_p/P_b > 13/2$, this percentage exceeds $100\%$. This indicates that the ring structure can be replenished with the material from
the planetesimal disk.

\begin{figure}
\centering \includegraphics[width=0.8\textwidth]{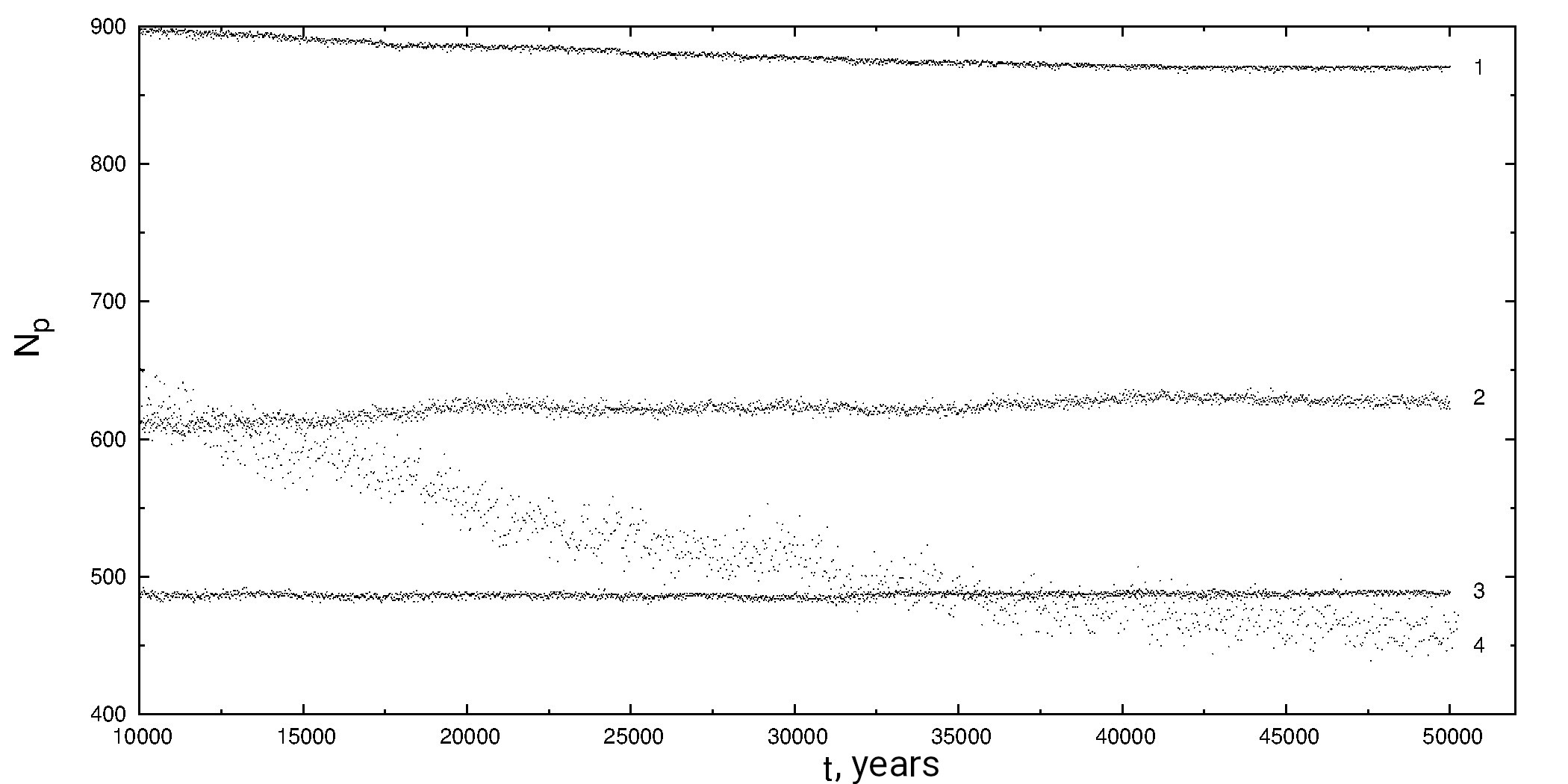}
\caption{The number of planetesimals in the coorbital structure as a function of time (in years). The parameters of the models are given in the text.}
\label{fig9}
\end{figure}

Figure~\ref{fig9} shows the rate of change in the amount of matter in the coorbital structure for four models. The model with the parameters $\mu_b = 0.5, m_p = 10M_J$, and $P_p/P_b = 8/1$ has the least stable coorbital ring: it retains less than $97.06\%$ of matter (curve 1). The most stable ring ($100.33\%$) is in the model with $\mu_b = 0.5, m_p = 1M_J$,
$P_p/P_b = 11/2$ (curve 3), while in the model with $\mu_b = 0.5, m_p = 0.3M_J$, and $P_p/P_b = 8/1$ the amount of matter in the ring grows more rapidly ($102.95\%$) than in the other cases (curve 2). For the models with a single central star, the percentage of surviving particles in the
coorbital structure does not exceed $85\%$ and decreases with increasing $\mu_p$ (curve 4).

\newpage
\section*{CONCLUSIONS}

The study of ring-like structures that are coorbital with planets, based on the models with different parameters, has shown that the dynamics of matter within the ring's boundaries differs considerably in the case of binary and single star systems. The binary systems feature the orbits of the horseshoe type only, while the single star systems are dominated by the tadpole orbits, and the number of the latter increases with increasing mass of the planet. In turn, an increase in the planet's mass in the binary system leads to an increase in the width of the coorbital structure.

The empirical dependence of the width of the coorbital structure on the semimajor axis of the planet and the mass ratio of the planet and binary or single star is obtained from the analysis of $27$ models. In both cases, it turns out to be weaker than the dependence of the Hill radius on the same parameters ($\Delta a_{Hill} \propto 0.693\mu^{1/3} a_p$) in the restricted three-body problem~\citep[see][]{1999ssd..book.....M}. An increase in the ring's width with increasing ratio of the planet's mass to the mass of the central objects is higher in the case of a binary star. The dependence of the coorbital ring's width on the mass ratio of the binary star components is not found; however, the calculations have shown that the maximum surface density shifts toward the center of mass of the system with an increase in the parameter $\mu_p$.

Additionally, the study of coorbital rings using the models with widely varying parameters allowed us to confirm the previous results regarding the greater stability of such structures in binary star systems as compared to a single star~\citep{2016MNRAS.463L..22D}. That being said, the rate of planetesimal depletion is higher in the star’s vicinity and for massive planets, while in the case of remote and low-mass planets, the coorbital structure may be replenished with the material of the debris disk. The calculations have shown that the threshold mass ratio of the planet and binary star, at which the formation of the ring is still noticeable, corresponds to the value $\mu_p \sim 5 \times 10^{-5}$.

It should be noted that we have considered the case where the planet moves in the same plane with the binary system. The inclination of the planetary orbit relative to the binary star's orbit and the plane of the
disk can affect the results. However, our calculations have not revealed any significant changes in the results at a small inclination ($<5^\circ$). In~\citet{1997MNRAS.285..288L,2010AstL...36..808G}, it was shown that the motion of planets and low-mass companions, whose orbits were inclined to the disk and located close to the single star, distorted the inner boundary of the disk. For this reason, in the case of a moderate inclination ($<15^\circ$) of the planet to the protoplanetary
disk of the binary system, the coorbital ring will probably also be inclined with respect to the periphery of the disk. The formation of the coorbital structure at planetary orbit inclinations of more than $15^\circ$ is hardly possible, since, in this case, the planet resides outside the disk for the most of its orbital period. Additionally,
the question of existence of planets with such an inclination to the binary stars’ orbits remains open. All the circumbinary planets discovered so far move close to the plane of the central binary star. Additionally, the results of the numerical modeling of planetary evolution have shown that the planet eventually settles toward the disk's plane~\citep{2013MNRAS.431.1320X}.

The calculations have not considered the accretion of particles onto the stars and the planet. However, the analysis of the results has shown that particles can approach the planet and concentrate near its position, further moving synchronously with it. With time, the number of particles around the planet increases; this effect can lead to the growth of the planet at the planetesimal disk stage.

\section*{ACKNOWLEDGMENTS}
The author would like to thank I. I. Shevchenko for the valuable discussions of the results and important comments. The study is funded by the grant of the Russian Federation for Basic Research no 17-02-00028-a.

\bibliographystyle{rusnat} 
\bibliography{biblio}{}

\end{document}